\documentclass[review]{elsarticle}

\usepackage{lineno,hyperref,booktabs,multirow}
\usepackage{soul}
\usepackage[bottom]{footmisc}
\usepackage{xcolor}
\usepackage[]{todonotes}

\modulolinenumbers[5]

\begin{document}

\begin{frontmatter}

\title{The Unlocking of High-Pressure Science with
Broadband Neutron Spectroscopy
at the ISIS Pulsed Neutron \& Muon Source}




\author{
Jeff Armstrong,\textit{$^{a,\ast}$}
Xiao Wang,\textit{$^{a}$}
and Felix Fernandez-Alonso{$^{b,c,d,e,\ast}$}}

\cortext[]{jeff.armstrong@stfc.ac.uk (Jeff Armstrong);
felix.fernandez@ehu.eus (Felix Fernandez-Alonso)}

\address[]{~ISIS Facility, Rutherford Appleton Laboratory, Chilton, Didcot, Oxfordshire OX11 0QX, United Kingdom}
\address[]{~Materials Physics Center, CSIC-UPV/EHU, Paseo Manuel de Lardizabal 5, 20018 Donostia - San Sebastian, Spain}
\address[]{~Donostia International Physics Center (DIPC), Paseo Manuel de Lardizabal 4,
20018 Donostia - San Sebastian, Spain}
\address[]{~Department of Physics and Astronomy, University College London, Gower Street, London WC1E 6BT, United Kingdom}
\address[]{~IKERBASQUE, Basque Foundation for Science, Plaza Euskadi 5, 48009 Bilbao, Spain}

\begin{abstract}
Following significant instrument upgrades and parallel methodological
developments over the past decade,
the TOSCA neutron spectrometer
at the ISIS Pulsed Neutron \& Muon Source in the United Kingdom
has developed a rich and growing scientific community spanning a broad range of
non-traditional areas of
neutron science, including chemical catalysis, gas adsorption \& storage, and new materials
for energy and sustainability. High-pressure science,
however, has seen little to no representation to date owing to
previous limitations in capability.
Herein, we explore for the first time the viability of rapid high-pressure measurements
in the gigapascal regime,
capitalizing from the orders-of-magnitude increase in incident
flux afforded by a recent upgrade of the primary-beam path.
In particular,
we show that spectroscopic measurements up to pressures of $\sim$2~GPa over 
an unprecedented energy-transfer range are now possible within
the hour timescale. In addition, we have designed and commissioned a dedicated set of high-pressure vessels, with a view to foster and support the further growth and development of an entirely new user community on TOSCA.
\end{abstract}

\begin{keyword}
\texttt{
inelastic neutron scattering,
broadband neutron spectroscopy,
high pressure, gigapascal, TOSCA,
ISIS Pulsed Neutron \& Muon Source,
VESPA,
European Spallation Source}
\end{keyword}

\end{frontmatter}


\section*{Introduction}
Over the past couple of decades, Inelastic Neutron Scattering (INS) has
moved from being a relatively exotic technique primarily used by condensed-matter
physicists, to becoming a mature and much-needed addition in the modern multi-pronged approach to understanding new materials. The major benefits of INS compared with other spectroscopic techniques are widely understood and well documented \cite{nbook2013,nbook2015,nbook2017}. Its chief benefits are its highly penetrating nature, the absence of hard spectroscopic
selection rules, and a very high incoherent neutron-scattering cross-section for
the proton, providing extreme sensitivity to motions involving hydrogen atoms. In addition, the past decade has witnessed a very rapid increase in the quantity and quality of combined in-silico and INS studies, giving unprecedented clarity
to a range of chemical phenomena at the atomic scale \cite{druzbicki2021,Armstrong2020,Dymkowski2018}.
Some examples include the understanding of phase transitions \cite{RosuFinsen2020,Kieslich2018,Butler2019}, quantifying the role of vibrations to the kinetics in catalytic processes \cite{Armstrong2021} and  gas absorption in porous media \cite{Savage2016}.

\begin{figure}[h]
    \centering
    \includegraphics[width=1.0\textwidth]{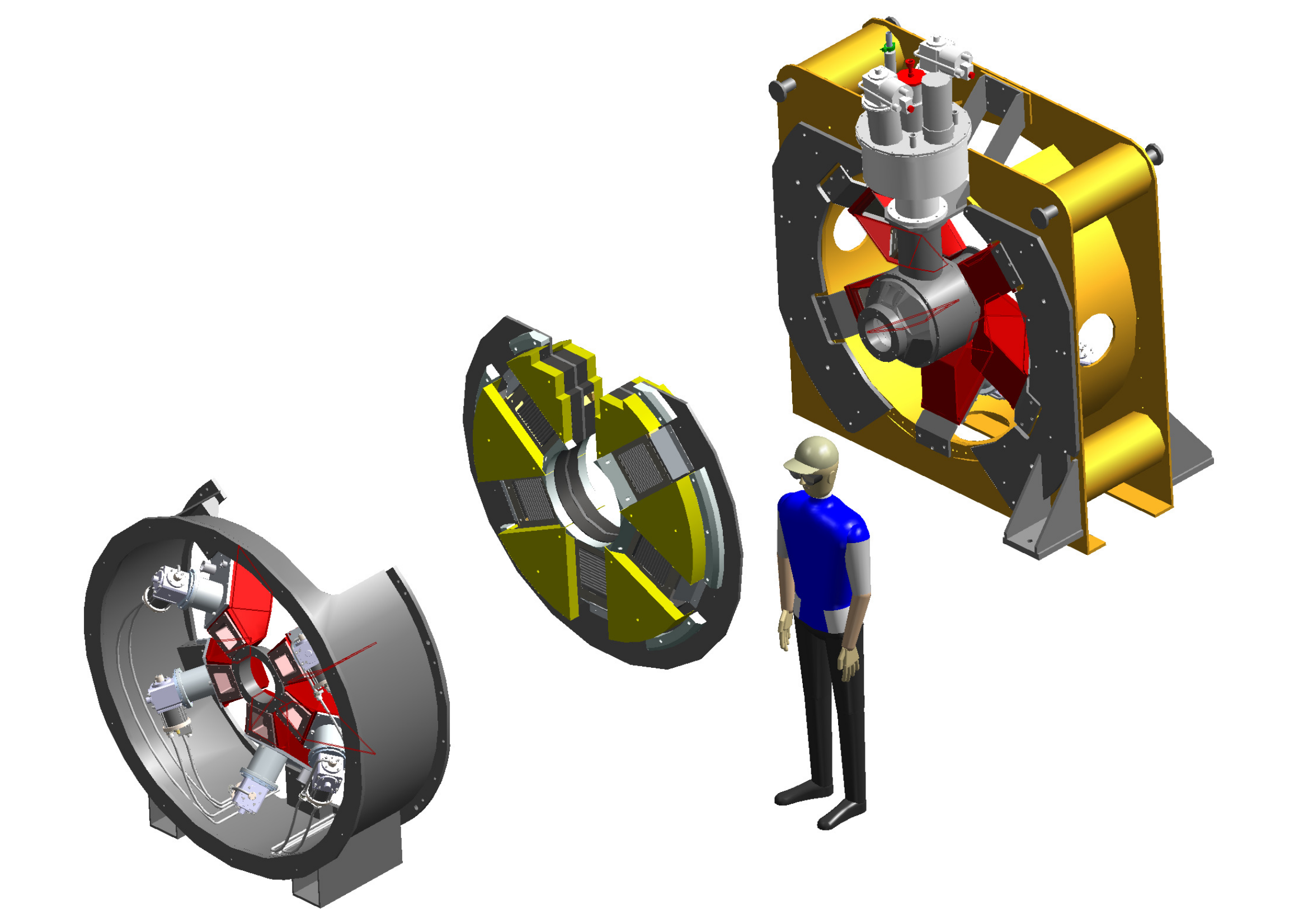}
    \caption{Schematic drawing of the TOSCA secondary spectrometer, where the front and back analysers (red boxes) have been split from the detectors (yellow wedges)
    in order to improve the visibility of the various components.}
    \label{toscageom}
\end{figure}

The TOSCA spectrometer at the ISIS Pulsed Neutron \& Muon Source in the United Kingdom
is a prime example of the above paradigm shifts, with a strong science programme in the chemical sciences.
Notwithstanding,
there have been very few High-Pressure (HP) studies performed on TOSCA to date, given the additional
difficulties relative to crystallographic studies~\cite{Guthrie2017,Klotz2012}.
In fact, the few examples conducted
with relative success were performed over twenty years ago on TFXA, TOSCA's predecessor. In particular, we note the work of Adams~\cite{phdthesisadams}, who designed a bespoke cadmium collimator to shape the incoming beam in order to match the sample geometry of a McWhan press, while also collimating the scattered beam to reduce noise from other sourcres of parasitic scattering.
This approach was necessary in order to use the rather bulky McWhan press, whose structural integrity is assured by the use of thick steel components. Despite some moderate success with this setup on TFXA~\cite{phdthesisadams,Belushkin1997,Adams1995}
as well as on other neighbouring instruments like IRIS~\cite{Sieber2006},
one major issue with such a cell is that the cooling times were typically of at least
15 hours to reach base temperature, severely curtailing the amount of time left for each measurement. Since these (rather heroic) initial and quite commendable
efforts, there have been significant improvements
to the original TFXA instrument
on its transition to the current incarnation of the TOSCA spectrometer.
These include additional detector banks positioned both in forward and back-scattering geometries and the addition of cadmium-slatted cooled beryllium filters to the final-energy
analysers~\cite{tosca2014} (see Figure~\ref{toscageom}).
More recently,
a sophisticated primary-flight path exploiting the latest advances in neutron-guide
technology have resulted in an order-of-magnitude increase in incident flux~\cite{Pinna2018}.
On a parallel front, the instrument is also equipped to conduct simultaneous
Raman measurements, a much-valued
and increasingly popular addition in order to increase the information
content of a given experimental campaign~\cite{Adams2009}.

\begin{figure}[h]
    \centering
    \includegraphics[width=0.8\textwidth]{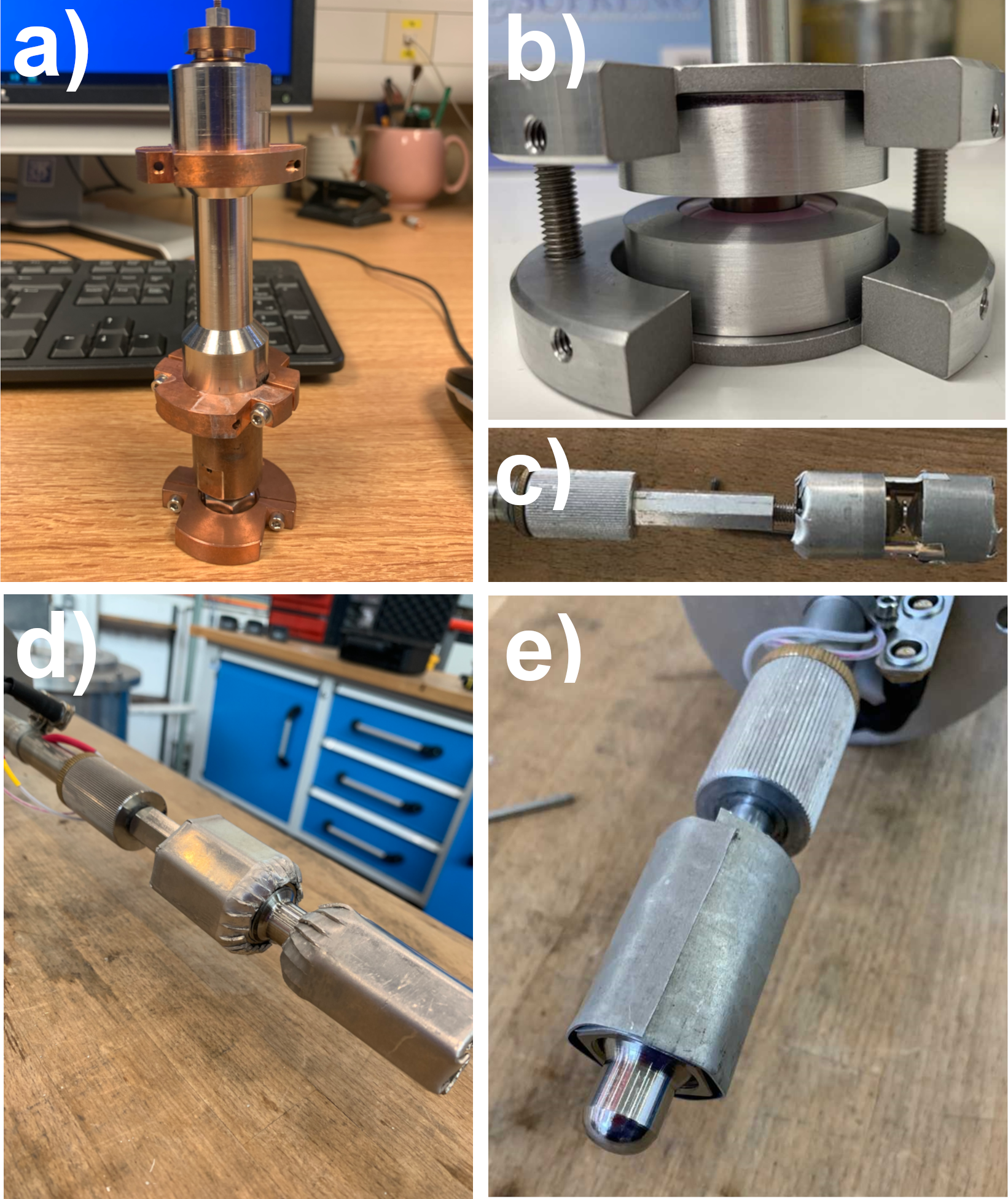}
    \caption{The range of HP cells discussed in the main text: a) a helium-driven piston cell; b) a clamped anvil cell with toughened zirconia-alumina anvil;  c) a sapphire anvil press with cadmium shielding; d) the ISIS clamp cell with cadmium shielding; e) the ILL clamp cell
    with cadmium shielding.}
    \label{differentcells}
\end{figure}

In the above context, we note that the highest gains
in flux (approaching two orders of magnitude) have been achieved
in the low energy-transfer range, which is of particular relevance to HP research. With these new capabilities, we pose the timely question as to whether HP measurements could finally become routine and time efficient. HP cells come in many varieties and span orders of magnitude in pressure.
Figure~\ref{differentcells} provides a visual overview of those of relevance to our subsequent
discussion below.
To facilitate the present discussion, we classify these different cell types into three categories.
At the lower end, there are the fluid-driven piston cells which typically
use a helium- or argon-gas intensifier to apply force to a piston which, in turn,
impinges upon the sample to deliver the requisite pressure.
These cells typically provide pressures of $\sim$0.5~GPa, allow sizeable sample volumes approaching the gram scale and their neutronic response is suitable to perform neutron-spectroscopic measurements,
in some cases simultaneously with high-resolution diffraction as recently demonstrated on
the neighbouring OSIRIS spectrometer~\cite{Fortes2017}.
At the high end of the spectrum, there are the likes of the diamond-anvil presses,
which rely on raw mechanical force being applied between the faces of two high-quality diamond anvils. These cells are capable of reaching $\sim$50-100~GPa, but require very small sample volumes ($\sim$10~mg).
In the middle ground between these two limits, we have simple clamped presses, which rely on a mechanical force being applied to a piston in contact with the sample and pressure medium. After the application of pressure, the piston us usually held in place with some form of locking mechanism
(or clamp). These presses are typically made of very-high-strength metal alloys with a
favourable neutronic response. Both the volumes and pressures for these cells are modest, achieving $\sim$2~GPa for a sample volume of $\sim$100-200~mg. In this situation,
the ability to acquire a useful INS signal involves a delicate balance between sample volume, the cell background and sample \& instrument geometries. Assessing these factors as well as how these
translate into new capabilities on TOSCA thus requires
a wide range of meaningful measurements for different sample sizes and cell geometries.

\section*{Assessment of current capabilities}

Sample sizes in a typical TOSCA experiment are on the few-gram scale ($\sim$2-4~g). In an initial
step, measurements on smaller samples using a standard aluminium sample cell
as a nominal background were conducted in order to understand the limits of detection of such samples
and thus establish a meaningful baseline for subsequent work with the various cell designs.
Compressed pellet samples were selected at representative sizes for an
anvil cell ($\sim$10~mg) and a press
cell ($\sim$100~mg). The chosen sample was that of 2,5-Diiodothiophene (2,5-D),
a commonly used and well-understood calibrant in
TOSCA-like neutron spectrometers~\cite{Pinna2018,Scatigno2022}.
Despite its modest hydrogen content, it does benefit from exhibiting
narrow and well-defined peaks and an uncongested spectrum, thus we expect it
to be a viable sample to assess smaller sample volumes. 

\begin{figure}[h]
    \centering
    \includegraphics[width=1.0\textwidth]{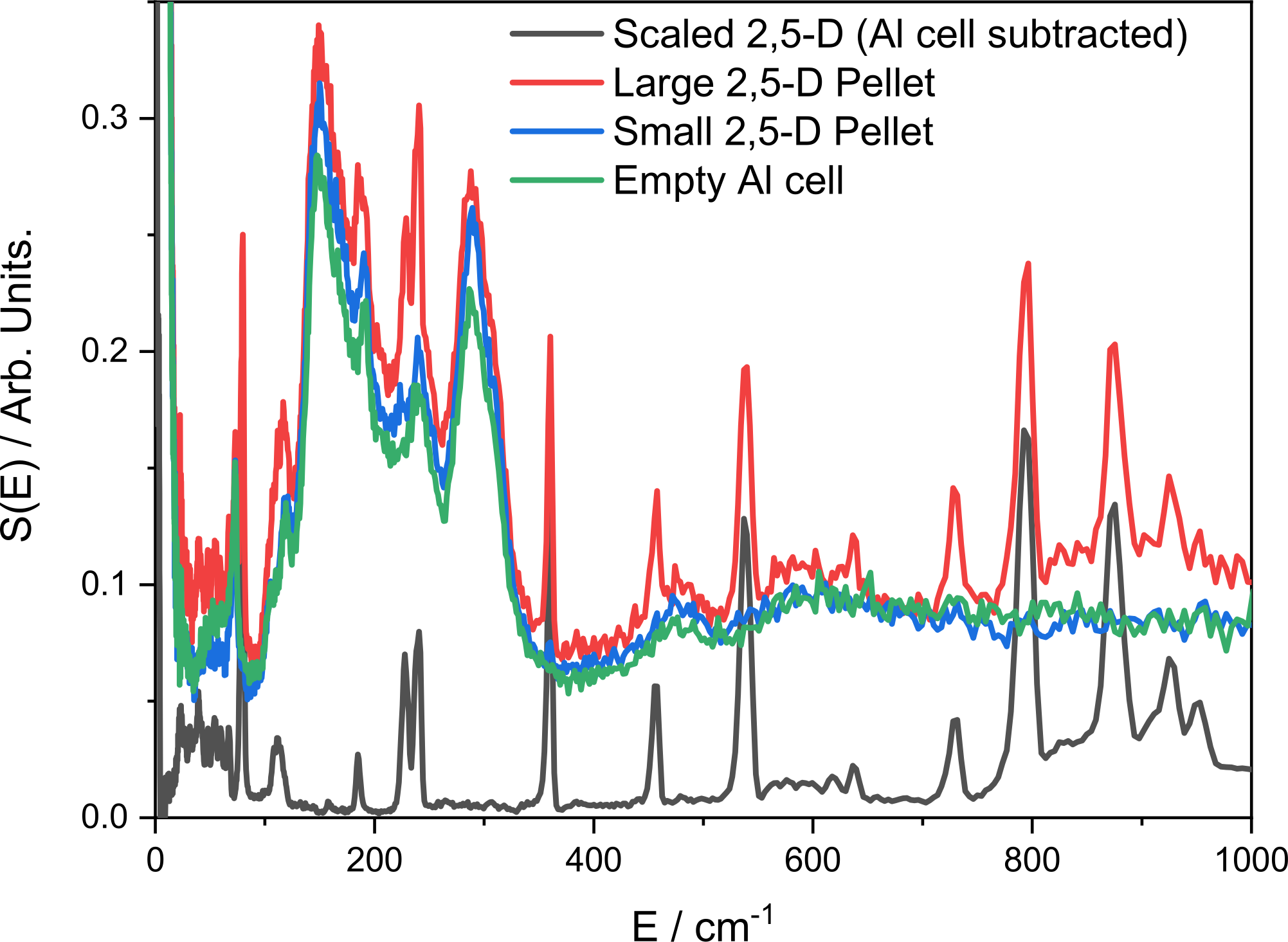}
    \caption{TOSCA spectra for different pellet sizes of 2,5-D as summarized in the figure legend.
    For further details, see the main text.}
    \label{pellets}
\end{figure}

\begin{figure}[h]
    \centering
    \includegraphics[width=1.0\textwidth]{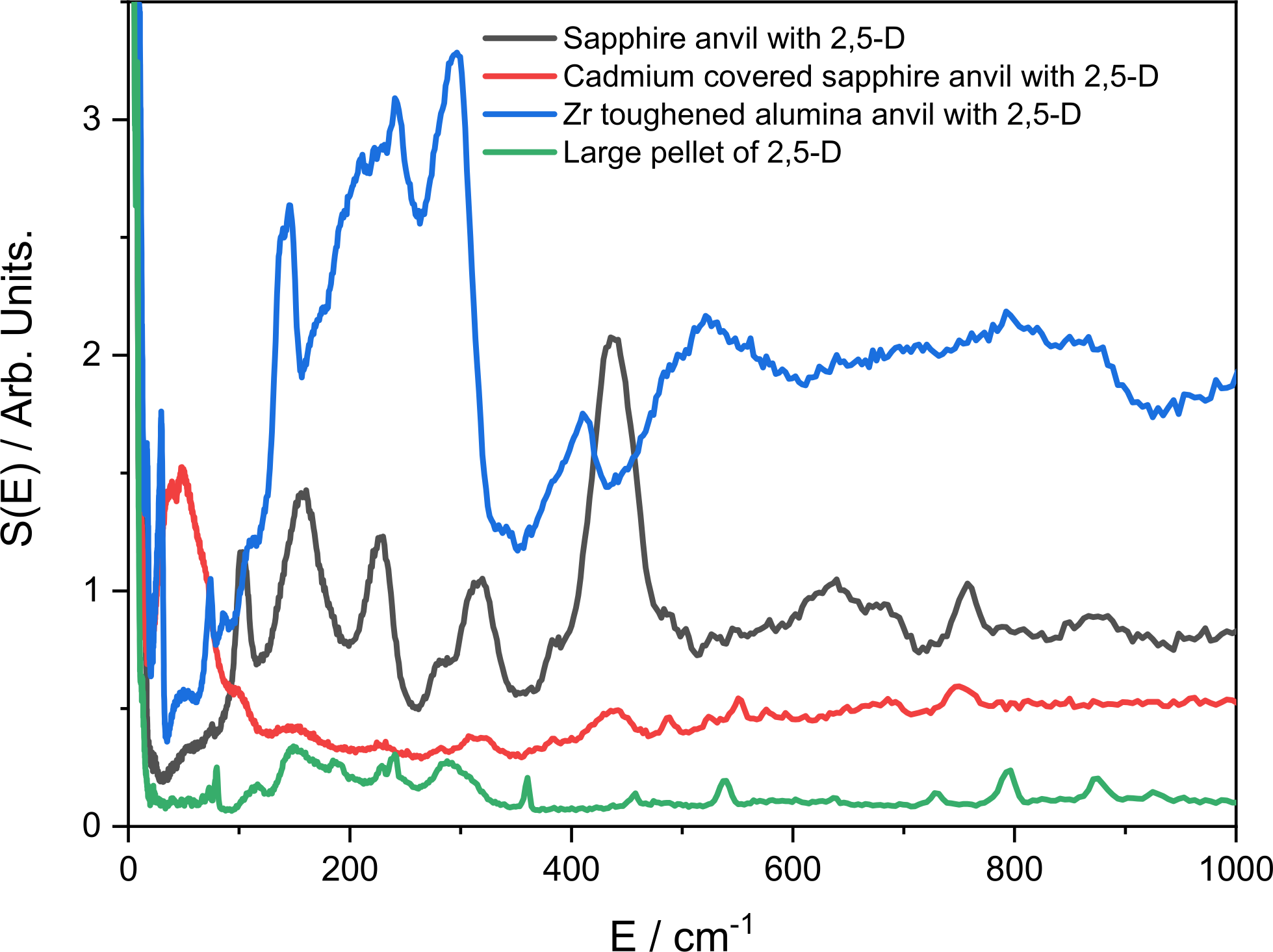}
    \caption{Spectra from the GPa cells containing a 2,5-D sample: the sapphire anvil press (red: with cadmium shielding; black: unshielded); zirconia-toughened anvil clamp cell (blue); and the large pellet as reference (green).}
    \label{sapphire}
\end{figure}

Experimental results for these measurements are presented in Fig.~\ref{pellets} alongside a scaled 3~g sample of 2,5-D to provide a suitable reference. From these data, it is clear that the signal from the smaller of the two pellets is too weak for detection with the given background of the standard
aluminium can, whereas the larger pellet is reasonably clear for all of the main peaks of the spectrum.
The region of the spectrum containing the aluminium-can contribution is relatively high in comparison to the sample peaks, and its rolling features at low energy transfers imply that one would require
a high-quality empty-cell spectrum in order to perform a high-quality subtraction. This sort of subtraction is rarely needed for standard TOSCA measurements, yet it is clear from these data that it will be essential for any HP measurement with similar background contributions from the empty vessel.
Despite the already-very-weak signals from the smaller pellet,
we also investigated the neutronic response of a cadmium-shielded sapphire anvil press
(see Fig.~\ref{differentcells}), which is capable of reaching higher ranges of pressure with a pellet size equivalent to the smaller of the two pellets. Figure~\ref{sapphire} shows both the shielded and unshielded spectra for the 2,5-D containing sapphire press.
Despite the shielding having a very large effect in reducing the background, the shielded spectra is still of the same order as a standard aluminium can, and thus the smaller pellet cannot be discerned from the background. It is clear then that in order for such a press to be used on TOSCA and similar
instruments there would either need to be a significantly lower background, or a large increase in flux on the instrument. The first of these two criteria is likely to remain a general limitation of all such anvil presses capable of these higher pressures. The second criterion could be fulfilled with a
secondary-spectrometer upgrade on TOSCA, which currently does not take advantage of its full solid-angle coverage of detectors, or benefit from more advanced focusing techniques such as parabolic analysers.
The viability of such measurements with future upgrades of TOSCA will be discussed in
more detail in the final section.

\begin{figure}[h]
    \centering
    \includegraphics[width=1.0\textwidth]{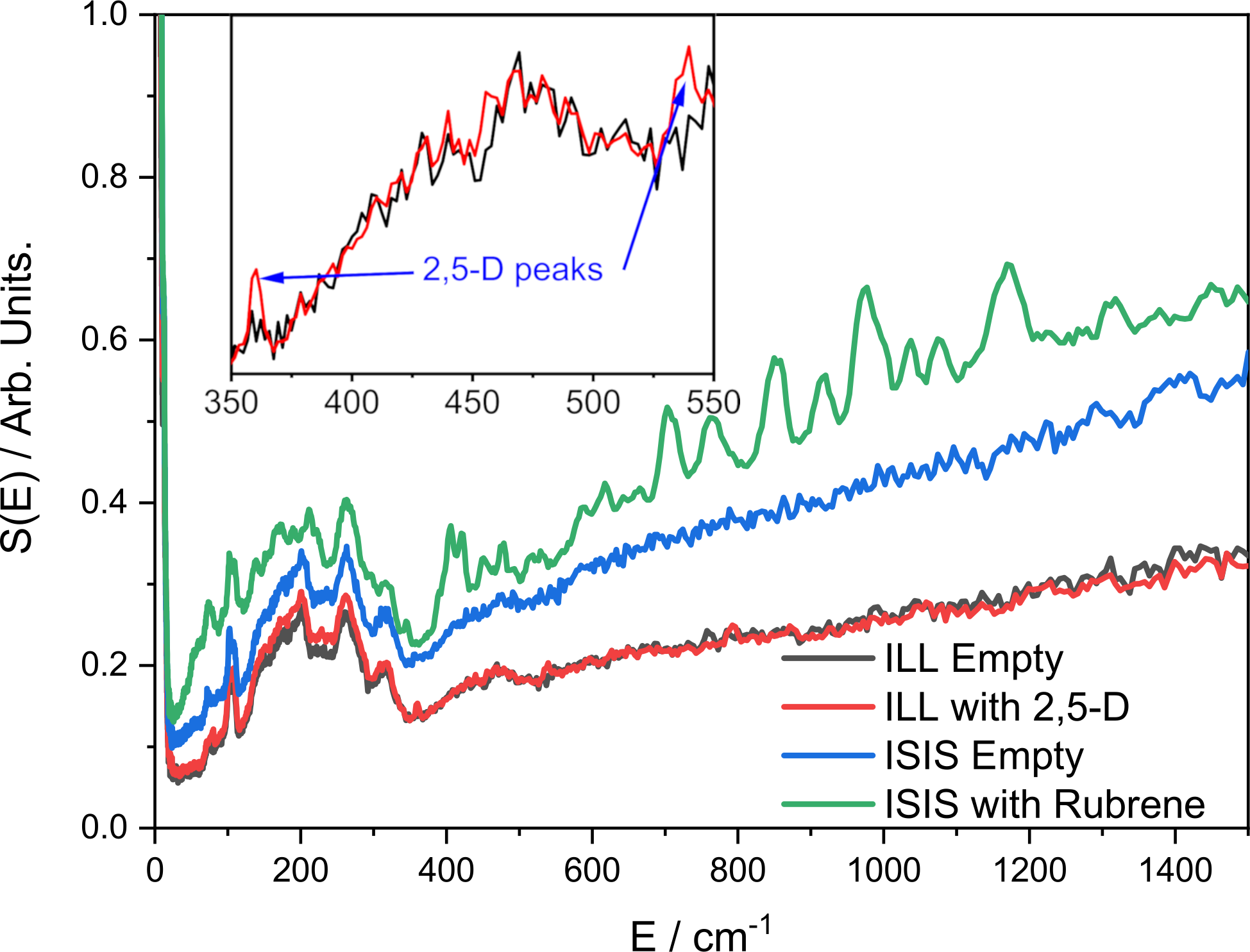}
    \caption{TOSCA spectra of the ILL and ISIS clamp cells with and without sample. The inset shows a zoomed in portion of the ILL data. For further details, see the main text.
    }
    \label{presscells}
\end{figure}

A similar style of anvil cell which allows a larger sample volume ($\sim$250~mg) at the expense of pressure ($\sim$3~GPa) was also tested. This press consisted of a steel fret and zirconia-toughened alumina anvil (see Fig.~\ref{differentcells}).
This geometry is not an optimal one as it does not take full advantage of the detector geometry of the
secondary spectrometer. This is due to the broadness of the support, combined with the shallow angle, which is much less than the position of the TOSCA detector banks at scattering angles of 45 and 135 degrees.
The scattered beam, therefore, has to pass through a very large portion of the support before reaching the detectors, thereby compromising the signal-to-background ratio.
These considerations are confirmed by the results shown in Fig.~\ref{sapphire}, showing a very large background which cannot be effectively reduced by cadmium shielding, as scattered neutrons would be also blocked by such a shielding.
With anvil cells seemingly unsuited to the current incarnation of TOSCA, we turn our attention to clamped cells. Although they achieve lower maximum pressures, they still outperform gas-driven cells
quite significantly (a factor of $\sim$4 increase in pressure) and allow us to enter the GPa regime.
Two cell designs were investigated, one which was designed for the ILL spectrometer IN1-LAGRANGE (see Fig.\ref{differentcells}-e)~\cite{Ivanov2014},
and another which is an existing ISIS cell used on a range of
instruments (see Fig.\ref{differentcells}-d). Neither cell was designed specifically
for TOSCA, thus we anticipate that their response
could still be improved significantly. LAGRANGES's detector array is a half-dome geometry, which is reflected in the cell design with a rounded bottom and broader upper section. This design is not optimal for TOSCA, as scattered neutrons from the top of the sample have to pass through a portion of the upper body in their way to the graphite analysers. Nonetheless, the more favourable sample volume (0.1~cm$^{3}$)
to achieve maximum working pressures of up to $\sim$2~GPa already make it a  
seemingly more viable option than the anvil cells. The ISIS cell can attain similar pressures of
$\sim$1.7~GPa with a significantly larger sample volume ($\sim$0.35~cm$^{3}$).
Figure~\ref{presscells} shows the results for both of these clamp cells. The first thing to notice is that the backgrounds for both cells was similar, and were both comparable to the standard aluminium cell, immediately giving an indication that a 2,5-D sample of $\sim$100~mg should be observable.
Indeed, the main peaks of the measured 100~mg sample were observable in the ILL cell as can be seen in the inset of Fig.~\ref{presscells}. It is clear, however, that the weaker features
and those contained within the most prominent features of the empty-cell spectrum
will be significantly harder to isolate with
confidence. With a similar overall neutronic response, the larger sample volume and more favourable
geometry of the ISIS cell would indicate that its performance could be significantly
better than the ILL design. As a test, we conducted measurements on the organic semiconductor rubrene.
As shown in Fig.~\ref{presscells}, spectral features associated with this material are readily
discernible from the experimental data without can subtraction after \textit{ca.} one hour of data collection, with publication-quality statistics in $\sim$2-3 hours. 
On TOSCA, we underline that 
this is the first time that measurements of this nature have been demonstrated
on such short timescales, meaning that such measurements could be performed routinely, thus
opening up a new research area on an already successful science programme on the instrument.
Such a clamped cell not only benefits from significantly higher pressures than gas-driven piston cells, but also in significantly simpler and faster setup times. Gas-driven high-pressure
setups require the operation of
bulky intensifier systems and substantial manpower to run and maintain them.
In addition, gas cells and their associated pipework require intricate leak
checking at different stages of assembly, and their overall safety
rating is substantially lower than that of a clamped cell.

\section*{Bespoke TOSCA design}

\begin{figure}[h]
    \centering
    \includegraphics[width=1.0\textwidth]{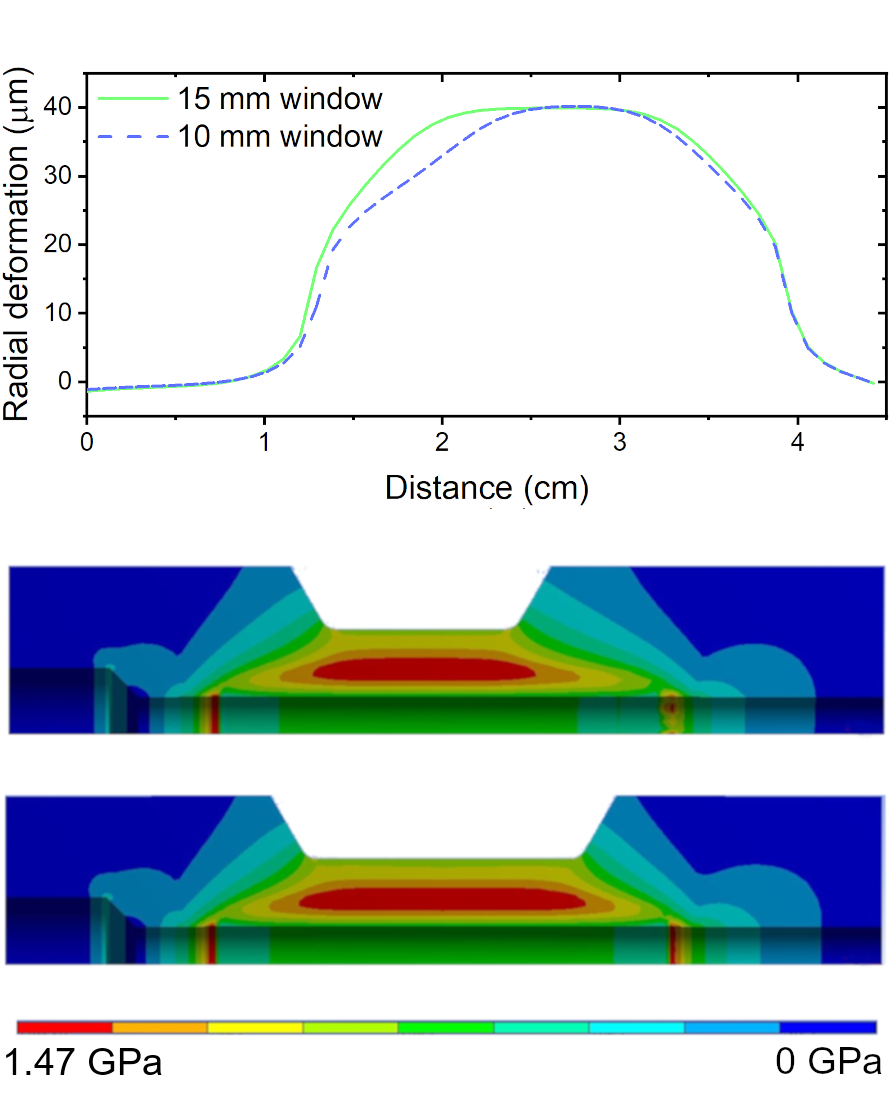}
    \caption{Upper panel: radial deformation of the cell as a function of the distance along each cell (15~mm window, solid green line;and 10~mm, dashed-blue line).
    Lower: stress cross-section for the cells with 10 and 15-mm windows.}
    \label{fecalcs}
\end{figure}

The results presented in the previous section clearly indicate that clamped cells constitute
the most suitable type of HP vessel to take TOSCA´s science programme into the GPa regime under realistic temporal scales. As these cells were not designed specifically for the instrument, there are further gains
in performance that deserve further thought and consideration.
To this end, a new bespoke clamped-cell design has been explored. For TOSCA, the existing ILL and ISIS cells both have a cylindrical sample space which is somewhat narrow in the radial dimension ($\sim$4~mm) and longer in the vertical dimension (8~mm for the ILL, and 24~mm for the ISIS press). The current TOSCA beam profile is 40x40~mm, thus an increase in either the radial or vertical dimension would result in a larger measured sample volume, and thus useful signal. The radial dimension is unlikely to provide a gain as an increase in radial sample volume requires a thickening of the walls of the cell to maintain a given maximum working pressure, and thus proportionate increase in background. The effect of increasing the vertical dimension however is not as clear from an engineering perspective. A set of Finite Element (FE) models run using ANSYS to simulate the cell body subject to maximum internal pressure. Two FE models were performed with the same boundary conditions and loading criteria (2~GPa internal pressure). The material for the main bore was chosen
to be the so-called "NiCrAl Russian alloy." The maximum internal pressure of the bore was chosen to be below the plastic threshold, as tabulated for this alloy. Both calculations were equivalent except for the length of the central window, which was chosen to be 10~mm and 15~mm. In order for a larger window size to be a safe and viable option, we also require that this lengthening does not create a larger area of the cell under maximal stress or, equivalently, a larger deformation area. Figure~\ref{fecalcs} shows both the radial deformation profile along the cell for both window sizes. In addition, a cross-sectional plot of the
stress profile is shown. The deformation plot clearly shows that although the maximum deformation is the same for both lengths of cell, the length over which this deformation is spread out is significantly longer for the larger window size. The stress cross-section also shows that the area over which there is a maximum stress is significantly larger, thus representing a significantly higher probability of failure.

\begin{figure}[h]
    \centering
    \includegraphics[width=1.0\textwidth]{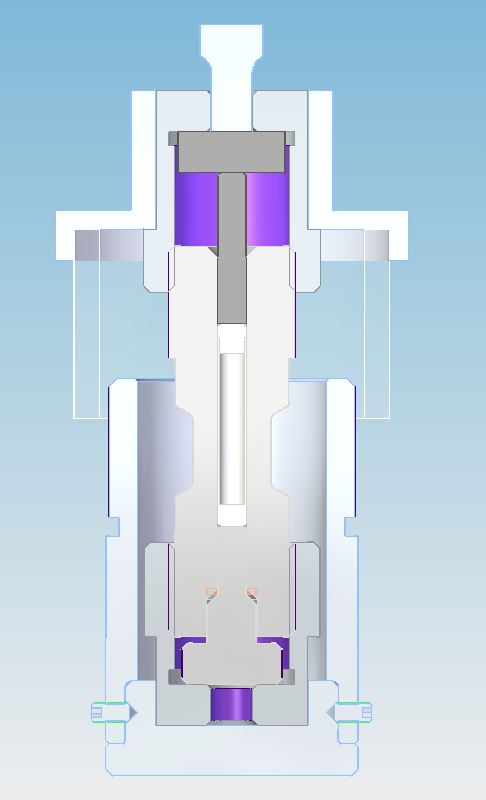}
    \caption{The bespoke TOSCA clamp cell, as described in more detail in the main text.}
    \label{newcell}
\end{figure}

This increased chance of failure applies to both a given use during an experiment as well as a shorter overall lifetime of the cell.
This consideration indicates that the narrower window constitutes
the best compromise between neutronic performance and risk
at the present time. Increased safety during operation was also deemed to be an important design
parameter. Clamp cells are generally regarded as moderately safe relative to other competing cell designs,
yet repeated use can led to catastrophic bursts and the violent ejection of sharp metal fragments.
To minimize risk, our bespoke design includes an in-situ shielding mechanism, in the form of a lower metal casing and an upper section with a transparent window to ensure full visibility of the cell at all
times. Figure~\ref{newcell} provides a visual summary of the design, including these additional features.
In addition, the previous ISIS design included two pistons which press from either direction. This
choice complicates significantly the initial pressurisation phase, requiring incremental tightening on both sides to ensure the sample is positioned in the central window. A single piston has therefore been used for the new design as this does not effect the maximum pressure, and simplifies the setup.
Finally, the body of the cell will be machined to contain flat sections which allow a wrench to be attached to lock the cell in place.

\section*{Conclusions and outlook}

We have shown that the increase in flux arising from the recent upgrade of the TOSCA primary spectrometer
continues to open new and unexplored vistas for its science programme,
with measurement times on the timescale of the hour that
enable broadband neutron spectroscopy at high pressures beyond the "GPa barrier."  
From the viewpoint of resources, we also underline that the infrastructure and manpower requirements of clamp cells are significantly lower than those of gas-intensifier alternatives. And the latter are also limited to operate below the GPa. Gas-driven pressure cells benefit from the flexibility of being able to readily scan a range of pressures and, thus, we anticipate that they will still remain an important option at the lower-end of the scale.
Clamped press cells, however, open up new research areas, especially in the case of phase transformations, where a single comparison above and below a transition
is typically required.
This work has also highlighted that the ability to reach even-higher pressures will be dependent on upgrades of the current TOSCA spectrometer. Namely, the upgrade of its secondary spectrometer would provide an increased flux, thus increasing the sensitivity to smaller sample volumes that are typical of higher-pressure cells. Such an upgrade employing curved graphite analysers with increased mosaicity
has been proposed~\cite{Zanetti2020} and design work
has already been initiated. This upgrade is estimated to give a gain of approximately one order of magnitude and, so, HP capabilities could be increased quite significantly. In addition to this, TOSCA benefits from a high spectral resolution that is relatively insensitive to the divergence of the
incoming beam. Thus, another possibility to enhance HP capabilities would involve the addition of
a focusing snout, which would funnel the 40x40~mm$^2$ beam profile into a smaller area of an appropriate size for the small sample volumes of HP cells. For a realistic focusing of the current beam  down to, say, 10x10~mm$^2$, the gain would be as high as an additional factor of $\sim$16. High-pressure science
would certainly benefit from this additional boost, at relatively little cost.
All of these gain factors combined would lead to drastically different capabilities for which HP cells could be employed on TOSCA. Very similar, if not identical, considerations also apply to other broadband
neutron spectrometers sharing similar operating principles such as VESPA, currently under construction
at the European Spallation Source in Lund (Sweden)~\cite{Andersen2020,Zanetti2019}.

\section*{Acknowledgments/Declarations}
We would like to thank Dr Craig Bull and Dr Christopher Ridley for providing use of two of the studied anvil cells and for useful discussions. We would also like to thank Alexandre Ivanov for similarly providing the ILL clamp cell and equally insightful conversations.
We also thank UK Research \& Innovation for financial support and the
provision of beamtime on the TOSCA spectrometer.
Financial support from the Spanish Ministry of Science and Innovation (Grant PID2020-114506GB-I00 funded by MCIN/AEI/10.13039/501100011033) and
the Basque Government (Grant PIBA-2021-0026 and IKUR Strategy)
is gratefully acknowledged. We would like to declare no competing interests.


\bibliography{bibl}

\end{document}